%% file: iclr2027_conference.tex
\documentclass{article} 
\usepackage{iclr2027_conference,times}
\iclrfinalcopy

\input{math_commands.tex}

\usepackage{hyperref}
\usepackage{url}
\usepackage{graphicx}
\usepackage{subfig}
\usepackage{amssymb}

\title{Achieving Robust Channel Estimation Neural Networks on Doppler effects by Designed Training Data}

\author{Dianxin Luan, John Thompson \\
Institute for Imaging, Data and Communications\\
University of Edinburgh\\
Edinburgh, EH9 3JL, United Kingdom
}

\begin{document}

\maketitle
\pagestyle{plain}
\thispagestyle{plain}

\begin{abstract}
Channel estimation is crucial in wireless communication systems. Many existing works train and test their models using the same or similar channel. However, practical channels are infinitely diverse and time-variant while data-driven algorithms often degrade on new and previously unseen channels. This prohibits the practical implementation of neural networks for channel estimation, particularly in high speed scenarios characterized by rapidly channel profile changes and very high Doppler shifts. Online training can fine-tune models to adapt to new channels, but requires high computation resource and incurs significant latency for physical-layer devices. Moreover, online training cannot exhaustively cover all possible channels because there are infinitely many possible channels. Therefore, we investigate the ability of neural networks to generalize to wireless channels. This paper proposes design criteria to generate synthetic training data, which guaranty that the resulting networks perform robustly across wireless channels, especially for very high Doppler shifts and different Doppler spectra. It ensures that offline-trained neural networks require no prior information or online training to precisely estimate the new channels. To demonstrate general applicability, we use neural networks with different levels of complexity to show that the generalization achieved appears to be independent of architecture. From simulation results, neural networks achieve a certain performance across wireless channels with different unseen power delay profiles (PDP) and Doppler shifts. 
\end{abstract} 

\section{Introduction}
For sixth-generation (6G) wireless communications and beyond, orthogonal frequency division multiplexing (OFDM) is expected to continue as the baseband modulator \cite{wang2023road}. Therefore, acquiring precise channel state information (CSI) is essential for reliable links that provide high data rates. Channel estimation refers to the process of estimating channel state information of the propagation environments between the transmitter and the receiver end. The CSI estimate is subsequently used to mitigate channel distortions and recover the transmitted data. Least-squares (LS) and minimum mean squared error (MMSE) methods are widely investigated for channel estimation (CE) \cite{van1995channel, edfors1998ofdm, li1998robust, srivastava2004robust}. However, conventional approaches struggle to meet the increasingly stringent data-rate and reliability system requirements, motivating the exploration of machine learning approaches for overall system optimization \cite{ye2017power, kim2018communication, kim2018deepcode, neumann2018learning, kim2020deepcode, baur2024leveraging, allaparapu2026precoder, luan2026robust}. Unlike conventional methods that aim for closed-form mathematical expressions, neural network approaches learn nonlinear mappings directly from the data and often achieve locally optimal but improved performance, such as ChannelNet \cite{soltani2019deep}. With the development of the attention mechanism, \cite{luan2022attention, luan2023channelformer} proposed an encoder-decoder architecture, called Channelformer, to utilize the transformer encoder \cite{vaswani2017attention} to focus on the critical subset of input LS elements. However, data-driven algorithms often degrade on new channels. This instability prohibits the real-world implementation of neural networks for time-variant channels. To adapt to channel variations while mitigating catastrophic forgetting, \cite{kong2025representation} proposed a continuous learning-based channel estimation solution that preserves previously learned knowledge through several subnetworks. A VAE-assisted unsupervised out-of-distribution (OOD) detector identifies unseen channels, after which a new subnetwork is rapidly adapted by using only a few online samples. To fine-tune neural channel estimator under varying channel conditions, \cite{wang2026channelmae} proposed a two-branch masked autoencoder that enables label-free online training through self-supervised reconstruction of randomly masked resource elements. \cite{wang2025fedpda} proposed a federated collaborative learning solution with pruning for neural receivers while mitigating catastrophic forgetting. 
\subsection{Problem statement}
Although online training methods can fine-tune offline-trained neural networks to adapt to new channels, they have several critical drawbacks as follows: 

1) Latency and computational consumption: Collecting training samples and parameters updating will result in significant and unavoidable latency. For example, the slot length is between 0.0625 milliseconds and 1 millisecond as defined in the 3GPP TS 38.211 document and the receiver needs at least 62.5-1,000 milliseconds to collect each 1,000 examples. Moreover, the computational cost and latency of the training process are typically high, so it is difficult for battery-powered terminals to update neural networks promptly, especially in highly dynamic environments. Moreover, online training cannot exhaustively cover all wireless channels as the set of possible channels is infinite. 

2) Interpolation capability and performance loss: The superior performance achieved by channel estimation neural networks is primarily down to their interpolation capability for OFDM. However, this good performance cannot be retained through online training because training requires the ground-truth channel matrix for the entire data packet, including the ground-truth channel matrix at both the vacant pilot positions (frequency-domain interpolation) and the data symbols (time-domain interpolation). The complete and ground-truth channel matrix of the environment channel is unavailable in practical systems, which indicates the loss of this superior performance. As shown in Fig.7a and Fig.7b of \cite{luan2025achieving}, it effectively addresses the generalization of neural networks to different channel power delay profiles (PDP), i.e. retaining frequency-domain interpolation capability on unseen channels. For time-domain interpolation, it is challenging because statistical models are inaccurate, cumbersome and difficult to handle \cite{nissel2018doubly}, and the basis expansion model \cite{borah1999frequency} is used for time-variant channels. Therefore, offline-trained neural networks are also expected to generalize well to different Doppler shifts and Doppler spectra, to retain time-domain interpolation capability on unseen channels. 
\subsection{Main contribution and notion}
The key contributions are summarized as follows: 
\begin{itemize}

    \item This paper proposes design criteria to generate synthetic training data which guarantees that trained neural networks generalize well to different Doppler spectra and Doppler shifts. Regardless of the channels which neural networks operate on, offline-trained neural networks achieve stable performance without any knowledge of this channel at the design time. This provides robust generalization to previously unseen channels. 

    \item Ground-truth and complete channel matrices can be still used as labels because the training channel is synthetically generated. By being trained with the complete channel matrix, neural networks retain precise interpolation capabilities in both frequency and time domain for superior estimate performance when operating on new, previously unseen channels. 
   
\end{itemize}
\textit{Notations}: In this paper, we use uppercase bold letters for matrices $\mathbf{X}$ and lowercase bold letters for vectors $\mathbf{x}$. The complex Gaussian distribution with mean $\mu$ and variance $\sigma^2$ is written as $\mathcal{CN}(\mu, \ \sigma^{2})$. The conjugate transpose and inverse of matrix $\mathbf{X}$ are represented by $\left(\mathbf{X}\right)^{H}$ and $\left(\mathbf{X}\right)^{-1}$. The function $\mathrm{Tr}\left\{\cdot\right\}$ denotes the trace of a matrix and $\left\Arrowvert\ \cdot\right\Arrowvert_{F}^{2}$ denotes the Frobenius norm of the matrix. $\mathbf{I}_N$ represents an identity matrix of size $N \times N$, $\mathbb{E}\{\cdot\}$ is the expectation operator, and $\mathrm{diag}\left\{\cdot\right\}$ is the diagonal operator that converts a vector to a diagonal matrix. For scalars, the conjugate operation is denoted by $(\cdot)^{*}$. $\otimes$ denotes the time-varying convolution operator. $\left\|\mathbf \cdot \right\|_1$ denotes the $\ell_1$ norm. The operator $\circ$ represents the Hadamard product. 
\section{System architecture and propagation channels}
\label{System architecture and propagation channels}
This paper considers an OFDM cellular system with a subcarrier spacing of $f_{space}$ kHz. The source signal is processed by a Quadrature Phase Shift Keying (QPSK) modulator. The QPSK symbols are assigned to the data subcarriers. The pilot signals are known to the receiver, and the vacant pilot subcarriers are set to 0. Each slot consists of $N_f$ subcarriers and $N_s$ OFDM symbols. By following the 5G NR specifications in 3GPP TS 38.211, a single-symbol demodulation reference signal (DM-RS) pattern with three additional positions is deployed. The 2\textsuperscript{nd}, 5\textsuperscript{th}, 8\textsuperscript{th} and 11\textsuperscript{th} OFDM symbols are reserved for pilots ($N_{pilot} = 4$). For each pilot OFDM symbol, the second subcarriers of each indices of $L_s = 2$ subcarriers is reserved for pilot subcarriers. The inverse fast Fourier transform (IFFT) converts the slot signal from the frequency domain to the time domain. Then the Cyclic-Prefix (CP, of length $L_{\mathrm{CP}}$ samples) is added to the front of each symbol to eliminate intersymbol interference (ISI). The channel model is a multipath channel with $M$ paths with impulse response of 
\begin{equation}
h(\tau, t) = \sum_{m=0}^{M-1} a_me^{j2\pi f_{d, m}t}\delta(\tau-\tau_mT_s), 
\label{impulse response}
\end{equation}
where $a_m \sim \mathcal{CN}(0,A_m)$ is the path gain of the \(m\)th path, $A_m = \mathbb{E}\left\{\left\|a_m\right\|_1^2\right\}$ is the average path power, $\tau_m$ is the corresponding delay normalized by the sampling period $T_{s}$, $f_{d, m}$ is the corresponding Doppler shift and $f_{d}$ is the maximum Doppler shift. The time-varying channel coefficient of the \(m\)th path is denoted by $\gamma_m(t)=a_m e^{j2\pi f_{d,m}t}$. Conditioned on the Doppler shift, $\gamma_m(t)$ is a complex Gaussian process. Doppler shifts $\{f_{d,m}\}_{m=0}^{M-1}$ are independently sampled from the normalized Doppler spectrum $S(f)$ for channel realizations ($\forall m, \left\|f_{d, m}\right\|_1 \leq f_d$) and are independent of path gains. Conventionally, the Doppler spectrum $S(f)$ of wireless channels is characterized by the Jakes model, which has the theoretical formula of 
\begin{equation}
    S(f)=\frac{1}{\pi f_{d}\sqrt{1-\left(f/f_{d}\right)^2}}, 
\qquad \left\|f\right\|_1<f_{d}, 
\end{equation}
and $S(f)=0$ otherwise. Therefore, the $n$th sampled time-domain $\mathbf{h}(n; t)$ is 
\begin{equation}
    \mathbf{h}(n; t) = \sum_{m=0}^{M-1} a_me^{j2\pi f_{d, m}t}e^{-j\pi\frac{n+(N_f-1)\tau_m}{N_f}}\frac{\sin(\pi\tau_m)}{\sin(\frac{\pi}{N_f}(\tau_m-n))}. 
\label{channel tap}
\end{equation}
The sampled PDP $\mathbf{p}(n) = \mathbb{E}\left(\mid \mathbf{h}(n)\mid^2\right)$, then the channel gain at the $k$th subcarrier and $l$th symbol is 
\begin{equation}
\mathbf{H}\left(k, l\right) = \frac{1}{\sqrt{N_f}}\sum_{m=0}^{M-1} a_me^{j2\pi f_{d, m}T_ol}e^{-j2\pi \frac{k\tau_m}{N_f}}. 
\label{Phase}
\end{equation}
for $k = 0, 1,..., N_f-1$ and $T_o$ is the symbol duration. Then $E\{\mathbf{H}_{k, l}\mathbf{H}_{k_{'}, l{'}}^{*}\} = r_t(l-l')r_f(k-k')$ where $r_f, r_t$ is the frequency and time correlation that are separable as shown in (\ref{r_t}). For Jakes model, $r_t(\Delta l) = J_0(2\pi f_d T_o\Delta l)$ is the zeroth-order Bessel function. The sampled signal $\mathbf{y} = \mathbf{h}\otimes \mathbf{x} + \mathbf{n}$, where $\mathbf{n}$ is additive white Gaussian noise (AWGN) and $\mathbf{x}$ is the transmitted signal. Each slot is assigned a new channel realization. After removing the CP, the receiver converts the time-domain data to the frequency-domain by the FFT operation. Assuming that $\forall{\tau_{m}} \leq L_{CP}$, the received signal at $k$th subcarrier and $l$th OFDM symbol, $\mathbf{Y}(k, l)$, is of the form: 
\begin{equation}
\mathbf{Y}(k,l) = \mathbf{H}_{k,k}(l)\mathbf{X}(k,l) + 
\sum_{\substack{q=0 q\neq k}}^{N_f-1}
\mathbf{H}_{k,q}(l)\mathbf{X}(q,l)
 + \mathbf{N}(k,l), 
\end{equation}
where $\mathbf{N} \sim \mathcal{CN}(\mathbf{0}, \ \sigma_{N}^{2}\mathbf{I}_{N_f})$, $\mathbf{H}_{k,q}(l)$ represents the coefficient for inter-carrier interference (ICI) due to Doppler shifts, and $\sigma_X^2 \triangleq E\{(N_fN_s)^{-1}\left\Arrowvert\mathbf{X}\right\Arrowvert_{F}^{2}\}$. For analytical tractability, ICI is neglected in the derivation of the proposed method and $\mathbf{H} \in \mathbb{C}^{{N_f}\times N_{s}}$ is the channel matrix, while simulations evaluate their effectiveness in the presence of ICI. The received pilot signals are then extracted to provide a reference for the channel estimate. The received signal is equalized by the channel estimate provided to recover the transmitted signal. 
\subsection{Propagation channels} 
\label{Propagation channels}
It is modeled as a Rayleigh fading channel operating at carrier frequency $f_{s}$ = 3.5GHz. The 3GPP and other candidate PDPs are given in Table.~\ref{customized channels} with low, medium, and high delay spread environments. 
\begin{table}[htbp]
\caption{Propagation channel profiles}
\begin{center}
\begin{tabular}{|c|c|}
\hline
\multicolumn{2}{|c|}{Two-path fading channel} \\
\hline
Path delay& [0 320] ns \\
\hline
Average path gain & [0.0 -3.0] dB \\
\hline
\multicolumn{2}{|c|}{Extended pedestrian A (EPA)} \\
\hline
Path delay& [0 30 70 90 110 190 410] ns \\
\hline
Average path gain & [0 -1 -2 -3 -8 -17.2 -20.8] dB \\
\hline
\multicolumn{2}{|c|}{Extended vehicular A (EVA)} \\
\hline
Path delay& [0 30 150 310 370 710 1090 1730 2510] ns \\
\hline
Average path gain & [0 -1.5 -1.4 -3.6 -0.6 -9.1 -7 -12 -16.9] dB \\
\hline
\multicolumn{2}{|c|}{Extended typical urban (ETU)} \\
\hline
Path delay& [0 50 120 200 230 500 1600 2300 5000] ns \\
\hline
Average path gain & [-1.0 -1.0 -1.0 0.0 0.0 0.0 -3.0 -5.0 -7.0] dB \\
\hline
\multicolumn{2}{|c|}{Defined channel 1 (DC1)} \\
\hline
Path delay& [0 50 100 200 400] ns \\
\hline
Average path gain & [0.0 -2.0 -4.0 -8.0 -16.0] dB \\
\hline
\multicolumn{2}{|c|}{Defined channel 2 (DC2)} \\
\hline
Path delay& [0 50 120 200 230 500 1600] ns \\
\hline
Average path gain & [-7.0 -3.0 -1.0 0.0 -2.0 -5.0 -7.0] dB \\
\hline
\multicolumn{2}{|c|}{Defined channel 3 (DC3)} \\
\hline
Path delay& [0 50 120 200 230 500 1600 2300 5000] ns \\
\hline
Average path gain & [0.0 -14.0 -15.0 -1.0 -5.0 -1 -15 -10 -15.0] dB \\
\hline
\end{tabular}
\label{customized channels}
\end{center}
\end{table}
\subsection{Channel estimation and methods}
\label{Channel estimation and methods}
\textbf{Channel estimation is to estimate the channel matrix $\mathbf{H}$. }To minimize the distance between the received signal and the channel-distorted signal, LS estimate at the $k$th subcarrier is 
\begin{equation}
    \mathbf{\hat{H}_{ls}}(k) = Y_{k}X_{k}^{-1}, 
\end{equation}
where $Y_{k}, X_{k}$ denotes the received and transmitted signals for the $k$th subcarrier. This paper applies bilinear interpolation on $\mathbf{\hat{H}_{ls}^{pilot}} \in \mathbb{C}^{\frac{N_f}{L_{s}}\times N_{pilot}}$ (at the pilot positions) to predict the complete channel matrix. To $\mathop{\arg\!\min_{\mathbf{H}}}\Arrowvert \mathbf{H}-\mathbf{H_{ls}}\Arrowvert^{2}_{F}$, MMSE estimate for OFDM symbols is computed by 
%
\begin{equation}
    \mathbf{\hat{H}_{MMSE}} = \mathbf{W}^{H}\mathbf{\hat{H}_{ls}}, W = \left(\mathbf{R_{HH}} + \left(\frac{\sigma_N^2}{\sigma_X^2}\right) \mathbf{I}_{N_f}\right)^{-1}\mathbf{R_{HH}}, 
\label{MMSE}
\end{equation}
where $\mathbf{R_{HH}} \in \mathbb{C}^{{N_f} \times {N_f}}$ is the channel autocorrelation matrix with $\mathbf{R_{HH}}(k_1, k_2) = E\{\mathbf{H}_{k_1}\mathbf{H}_{k_2}^{*}\}$ for $\forall{k_1, k_2 \in [0, N_f - 1]}$, and $E\{\mathbf{H}_{k, l}\mathbf{H}_{k_{'}, l{'}}^{*}\} = r_t(l-l')r_f(k-k')$ as shown in (\ref{Phase}) where $r_f, r_t$ is the frequency and time correlation. $\frac{\sigma_N^2}{\sigma_X^2}$ is the inverse signal-to-noise ratio (SNR) and $\sigma^2$ denotes the normalized power variance. We use the actual channel vectors for each pilot symbol $\mathbf{H_{c}} \in \mathbb{C}^{N_f}$ and the actual channel vectors for the pilot subcarriers $\mathbf{H_p} \in \mathbb{C}^{\frac{N_f}{L_{s}}}$ to compute the correlation matrix by $\mathbf{R_{H_{c}H_p}} = \mathbb{E}\left\{\mathbf{H_{c}}\mathbf{H_{p}^{H}}\right\}, \mathbf{R_{H_pH_p}} = \mathbb{E}\left\{\mathbf{H_{p}H_{p}^{H}}\right\}$. These matrices are then substituted into (\ref{MMSE}) to calculate $\mathbf{\hat{H}_{MMSE}^{pilot}} \in \mathbb{C}^{N_f}$ for each pilot symbol. Bilinear interpolation is applied to this MMSE estimate to predict the complete channel matrix estimate. By incorporating ground-truth channel matrix of the environment channel and LS estimate as input, the MMSE method outperforms the LS estimate, but the implementation of the MMSE method is noncausal in practice. 

Channel estimation neural networks take $\mathbf{\hat{H}_{ls}^{pilot}}$ as input to infer $\mathbf{\hat{H}} \in \mathbb{C}^{N_{f} \times N_{s}}$ as estimate for complete channel matrix. The real and imaginary parts of complex matrix are stored in the third dimension of both the real input and the real output. To demonstrate that the proposed method is effective across neural architectures, along with the Channelformer network, InterpolateNet in \cite{luan2021low} is also deployed as an example of low complexity estimator, with only 9,942 parameters. 
\section{Proposed design criteria for Doppler}
\label{Proposed design criteria for Doppler}
The detailed operation of neural networks is difficult to understand as mathematical interpretability is still lacking. Assuming that neural networks learn a mapping $g\left(\hat{\mathbf{H}}_{\mathrm{ls}}^{\mathrm{pilot}}\right)$ using MSE loss for training, they achieve the target $\min_{g} \mathbb{E}_{\mathbf{H},\hat{\mathbf{H}}_{\mathrm{ls}}^{\mathrm{pilot}}}
\left[\left\|\mathbf{H}-g(\hat{\mathbf{H}}_{\mathrm{ls}}^{\mathrm{pilot}})\right\|_{F}^{2}\right]$. For a given input $\hat{\mathbf{H}}_{ls}^{pilot}$, the training loss is 
\begin{equation}
\begin{aligned}
\mathrm{MSE}(\hat{\mathbf{H}}, \mathbf{H}) &=
\mathbb{E}\left[
\|\mathbf{H}-\hat{\mathbf{H}}\|_F^2
\mid \hat{\mathbf{H}}_{ls}^{pilot} 
\right], \\ 
\frac{\partial \mathrm{MSE}}{\partial \hat{\mathbf{H}}} &= 2\left( \hat{\mathbf{H}} - \mathbb{E}[\mathbf{H} \mid \hat{\mathbf{H}}_{ls}^{pilot}] \right), \nabla^2 \mathrm{MSE} = 2\mathbf{I} \succ 0. 
\end{aligned}
\end{equation}

Because the Hessian matrix is positive definite and this loss is strictly convex, the corresponding stationary point is the unique global minimizer, which provides the solution that 
\begin{equation}
    \hat{\mathbf{H}} = \mathbb{E}\left[ \mathbf{H} \mid \hat{\mathbf{H}}_{ls}^{pilot} \right], 
\end{equation}
this matches the definition of MMSE estimator $\hat{\mathbf{H}}_{\mathrm{MMSE}} = \mathbb{E}\left[ \mathbf{H} \mid \hat{\mathbf{H}}_{ls}^{pilot} \right]$. Therefore, when neural networks have sufficient capacity, training data and optimization, they always achieve $g(\hat{\mathbf{H}}_{ls}^{pilot}) = \hat{\mathbf{H}}_{\mathrm{MMSE}}$. Offline training facilitates this mapping by allowing both the training dataset and computational resources to be scaled to any desired size without imposing additional latency or memory consumption for online implementation. Therefore, by using MSE loss for training, the behavior of neural networks is similar to MMSE filters. 

As the $r_f$ design has been discussed in \cite{luan2025achieving}, this work only considers the $r_t$ mismatch on a matched PDP. An alternative expression for channel autocorrelation is $\mathbf{R_{HH}} = \mathbf{D}\mathbf{\Lambda}\mathbf{D}^{-1}, \mathbf{D}\mathbf{D}^{H}=\mathbf{I}_{N_f}$ where $\mathbf{D} \in \mathbb{C}^{{N_f} \times {N_f}}$ is the discrete Fourier transform (DFT) matrix and $\mathbf{\Lambda} = \mathrm{diag}\{\mathbf{p}\}, \mathbf{\Lambda} \in \mathbb{R}^{{N_f} \times {N_f}}$ \cite{li1998robust}. To investigate the time correlation among the slot, Doppler shifts are expressed using the normalized DTFT frequency ($[-f_d,f_d] \subseteq [-0.5,0.5]$). For the $i$th diagonal element of its eigenvalue the corresponding spectrum normalized by $\sigma^{2}$ is given by 
\begin{equation}
    S_{i}(f) = \frac{\mathbf{p}(i)}{\sigma^{2}}S_{t}(f) + 1. 
\label{one_side}
\end{equation}
\textbf{Under the anti-causal assumption}, spectral factorization is performed to (\ref{one_side}) for one-sided function $M_i(f)$ that satisfies $\left\|M_i(f)\right\|_1^{2} = \frac{\mathbf{p}(i)}{\sigma^{2}}S_{t}(f) + 1$. Then converting $M_i(f)$ back by the inverse Fourier transform to obtain $\Gamma_{i}$, we have its filter taps $\Gamma_{i}$ expressed as 
\begin{align}
    & M_i(f) = \sum_{n = 0}^{\infty} \Gamma_{i}(n)e^{-jn2\pi f}, \\ & \sum_{n = 0}^{\infty}\Gamma_i^{*}(n)e^{jn2\pi f}\sum_{n = 0}^{\infty}\Gamma_i(n)e^{-jn2\pi f} = \mathbf{p}(i)\frac{\sigma_X^2}{\sigma_N^2}S_t(f) + 1, 
\label{filter}
\end{align}
as shown in (\ref{M_i}) of Appendix~\ref{Appendix}. In this case, the corresponding anti-causal Wiener filter is given by 
\begin{equation}
    \mathbf{w}_{i}(f) = 1-\frac{1}{M_i(-f)\Gamma_i(0)}, 
\end{equation}
and then the weight matrix at the design time is given by 
\begin{equation}
\begin{aligned}
    \mathbf{W}^{D}(f) &= \mathbf{D}\mathbf{w}^{D}(f)\mathbf{\Lambda}^{\dagger}\mathbf{D}^{H}\mathbf{R}_{HH} \\ &= \mathbf{D}\mathbf{w}^{D}(f)\mathbf{\Lambda}^{\dagger}
   \mathbf{D}^{H}\mathbf{D}\mathbf{\Lambda}\mathbf{D}^{H} = \mathbf{D}\mathbf{w}^{D}(f)\mathbf{D}^{H} . 
\end{aligned}
\end{equation}
The corresponding $\mathbf{w}^{D}(f) \in \mathbb{C}^{N_f \times N_f}$ is a diagonal matrix with $i$th elements ${\mathbf{w}_{i}^{D}}(f)$ of 
\begin{equation}
    \mathbf{w}_{i}^{D}(f) = 1 - \frac{1}{M_{i}^{D}(-f)\Gamma_i^{D}(0)}, 
\label{w_i}
\end{equation}
and $S_{t}^{D}(f)$ is the Doppler spectrum of the designed channel. It has 
\begin{equation}
    \left\|M_{i}^{D}(f)\right\|_1^{2} = 1 + \frac{\mathbf{p}(i)}{\sigma^{2}}S_{t}^{D}(f). 
\label{M_i^D}
\end{equation}
Assuming the $S_{t}^{A}(f)$ is the Doppler spectrum of the environment channel, the mismatch error caused by the Doppler spectrum mismatch between the designed channel and the environment channel $\Delta \epsilon_t(S_t^{D}; S_t^{A})$ = $\frac{1}{N_f}\sum_{i=1}^{N_f}\left(\epsilon_{t, i}\!\left(S_{t}^{A};\,S_{t}^{D}\right)-\epsilon_{t, i}\!\left(S_{t}^{D};\,S_{t}^{D}\right)\right)$ is 
{\footnotesize
\begin{equation}
    \Delta \epsilon_t(S_t^{D}; S_t^{A}) = \frac{1}{N_f} \sum_{i=1}^{N_f} \mathbf{p}(i)\int_{-0.5}^{0.5}\left(S_{t}^{A}(f) - S_{t}^{D}(f)\right)\left\|1-\mathbf{w}_{i}^{D}(f)\right\|_1^2\,df. 
\label{error} 
\end{equation}}
From (\ref{Gamma}) shown in Appendix~\ref{Appendix}, ${\Gamma_i^{D}(0)}^{2} = \exp\left\{\int_{-0.5}^{0.5}\ln\left(1+\frac{\mathbf{p}(i)S_t^{D}(f)}{\sigma^{2}}\right)\,df \right\}$. With $S_t^D(f) = S_t^D(-f)$ and $\Gamma_i^D(0)>0$, we have 
\begin{equation}
    \left\|1-\mathbf{w}_i^D(f)\right\|_1^2 = \frac{\sigma^2} {\left(\Gamma_i^D(0)\right)^2 \left(\mathbf{p}(i)S_t^D(f)+\sigma^2\right)}. 
\end{equation}
Then this mismatch error is simplified to 
\begin{equation}
\begin{aligned}
\Delta \epsilon_t &= \frac{1}{N_f} \sum_{i=1}^{N_f}\frac{\sigma^{2}}{{\Gamma_i^{D}(0)}^{2}}\int_{-0.5}^{0.5}\frac{\mathbf{p}(i)\left(S_t^{A}(f)-S_t^{D}(f)\right)}{\mathbf{p}(i)S_t^{D}(f)+\sigma^{2}}
\,df \\
&= \frac{\sigma^{2}}{N_f} \sum_{i=1}^{N_f}\frac{1}{{\Gamma_i^{D}(0)}^{2}}\left(\int_{-0.5}^{0.5}\frac{\mathbf{p}(i) S_t^{A}(f)+\sigma^{2}}{\mathbf{p}(i)S_t^{D}(f)+\sigma^{2}}\,df-1\right).
\end{aligned}
\end{equation}
Except the matched Doppler spectrum, the Doppler spectrum of the designed channel is set to be the flat spectrum which has the theoretical formula of 
\begin{equation}
S(f)=
\begin{cases}
\dfrac{1}{2f_d}, & \left\|f\right\|_1\leq f_d, \\ 
0, & \text{otherwise}.
\end{cases}
\end{equation}
It satisfies $S_t(f) = S_t(-f)$ and when the boundary $f_d^{D}$ of the flat spectrum design exceeds $f_d^{A}$ of $S_{t}^{A}(f)$, it has 
\begin{equation}
\int_{-0.5}^{0.5}\frac{\mathbf{p}(i) S_t^{A}(f)+\sigma^{2}}{\mathbf{p}(i) S_t^{D}(f)+\sigma^{2}}\,df = 1, \Delta \epsilon_t = 0. 
\end{equation}

Therefore, by using the flat spectrum with a sufficiently large $f_d^{D}$ for design, the resulting filter should generalize well to any Doppler spectrum shape with a smaller Doppler shift. It should be noted that $S_t^{D}(f)$ should be symmetric while the environment channel does not necessarily to be symmetric. 

\textbf{For the noncausal assumption}, the noncausal corresponding coefficient is given by 
\begin{equation}
    \mathbf{w}_i(f) = \frac{\mathbf{p}(i) S_t(f)}{\mathbf{p}(i) S_t(f)+\sigma^{2}}.
\end{equation}
Assume flat spectrum is deployed as the designed Doppler spectrum and the environment has a Doppler spectrum of $S_t^{A}(f)$, the $\mathbf{w}_i^{D}(f)$ is given by 
\begin{equation}
    \mathbf{w}_i^{D}(f) = \frac{\mathbf{p}(i)S_t^{D}(f)}{\mathbf{p}(i)S_t^{D}(f)+\sigma^2} = 
    \begin{cases}
    \frac{\mathbf{p}(i)}{\mathbf{p}(i)+2\sigma^2f_d^{D}}, & \left\|f\right\|_1\leq f_d^{D}, \\ 
0, & \text{otherwise}.
\end{cases}. 
\end{equation}
This result is independent of Doppler shifts, and $\left\|1-\mathbf{w}_i^{D}(f)\right\|_1^2 = \left(\frac{2\sigma^2}{\mathbf{p}(i) f_{d}^{-1}+2\sigma^2}\right)^2$ is a constant across Doppler shifts. Therefore, each mismatch error under this flat spectrum design is provided by 
{\small 
\begin{equation}
    \Delta\epsilon_{t, i} = \mathbf{p}(i)\left(\frac{2\sigma^2}{\mathbf{p}(i) f_{d}^{-1}+2\sigma^2}\right)^2 \int_{-0.5}^{0.5}\left(S_t^{A}(f)-S_t^{D}(f)\right)\, df =0. 
\end{equation}}
For arbitrary normalized Doppler spectrum, it holds that $\int_{-0.5}^{0.5} S_t^{A}(f)\,\mathrm{d}f = \int_{-0.5}^{0.5} S_t^{D}(f)\,\mathrm{d}f = 1$. Therefore, $\Delta\epsilon_t = 0$ when the flat spectrum is deployed as $S_t^{D}$ on any Doppler spectrum shape with a smaller value of $f_d$. Both causal and noncausal processing are commonly adopted in neural channel estimators (typically to predict for the whole slot), where pilot symbols embedded in intermediate OFDM symbols are used for time interpolation to predict the complete channel matrix of the whole slot. For example, the channel estimation at earlier symbols may depend on pilot symbols from later symbols, such as the first data OFDM symbol. Moreover, it shows that the design with flat spectrum minimizes the worst-case MSE over all admissible Doppler spectra $S$ under the Doppler bandwidth constraint. As we have $\varepsilon(\mathbf{w},S^D) \geq \varepsilon(\mathbf{w}^D,S^D)$ and $\sup_{S^A \in S} \varepsilon(\mathbf{w}^D,S^A) = \varepsilon(\mathbf{w}^D,S^D)$, then the following holds for any $\mathbf{w}$ that 
\begin{align}
    &\sup_{S^A \in S} \varepsilon(\mathbf{w},S^A) \geq \sup_{S^A \in S} \varepsilon(\mathbf{w}^D,S^A), \\ 
    &\mathbf{w}^D \in \arg\min_{\mathbf{w}} \sup_{S^A \in S} \varepsilon(\mathbf{w},S^A). 
\end{align}
Therefore, the flat design not only achieves MSE invariance across all possible spectra but also minimizes the worst-case MSE, thereby establishing its minimax optimality. 

The benefit is, neural networks only require offline-training to perform well on unseen channels, to avoid additional latency and computational assumption and retain the superior performance. We only need to consider the inference complexity for online implementation because training is completely conducted offline. As regularization prevents overfitting and improves generalization slightly, the regularization term is set to 0 to remove the potential contribution. This is to show generalization is primarily due to the design criteria for training. 
\section{Simulations}
\label{Simulations}
Normalized MSE (NMSE) is a performance metric that evaluates the normalized distance between the ground-truth channel matrix and its estimate for all resource elements in the slot, defined as 
\begin{equation}
    \mathrm{MSE}\left(\mathbf{\hat{H}}, \mathbf{H}\right) = \frac{\mathbb{E}\left\{\left\Arrowvert\mathbf{\hat{H}} - \mathbf{H}\right\Arrowvert_{F}^{2}\right\}}{\mathbb{E}\left\{\left\Arrowvert\mathbf{H}\right\Arrowvert_{F}^{2}\right\}}. 
\end{equation}
The offline-training dataset consists of 300,000 independent samples (95\% for training and 5\% for validation), which use the CE channel in \cite{luan2025achieving} as the default PDP. It ensures robust generalization to different PDPs, i.e. frequency-domain Interpolation. The mobile speed is set to 350km/h (maximum Doppler shift $f_d$ of 1135Hz) with a fixed SNR of 20dB. This setting represents an intermediate high-mobility condition between the 300km/h and 500km/h HST test profiles specified in 3GPP TS 38.101-4 \citep{3gpp381014}. The deployed loss function is the MSE loss and the system configuration and training hyper-parameters are shown in Table.~\ref{system hyper-parameters}. It should be noted that this work only uses a very small subcarrier spacing $f_{space}$ of 30kHz for high Doppler shifts scenarios. To average out Monte Carlo effects, each sample of the simulation curves is tested with 5,000 independent channel realizations. 
\begin{table}[t]
\caption{Simulation hyper-parameters for the deployed baseband configuration and the hyper-parameters for offline-training procedure of both Channelformer and InterpolateNet. }
\begin{center}
\begin{tabular}{llll}
\multicolumn{2}{l}{\bf Offline-training hyper-parameters} & \multicolumn{2}{l}{\bf System configuration} \\ 
\hline 
Optimizer& Adam& $N_f$& 72\\
\hline
Maximum epoch& 100& $N_s$& 14\\
\hline
Initial learning rate (lr) & 0.01& $L_{\mathrm{CP}}$& 12\\
\hline
Drop period for lr& 20& $f_{c}$& 3.5GHz\\
\hline
Drop factor for lr& 0.2& $f_d$ & 1135Hz\\ 
\hline
Minibatch size& 64& $f_{space}$& 30kHz\\
\hline
L2 regularization& \textbf{0}& $T_{s}$& 4.629e-07s\\ 
\hline
\end{tabular}
\label{system hyper-parameters}
\end{center}
\end{table}
\subsection{Generalization to different Doppler spectrum and Doppler shifts}
\label{Doppler_MSE}
To show the generalization to different Doppler conditions, offline-trained neural networks are tested with previously unseen Doppler spectra and Doppler shifts. We only use the flat Doppler spectrum and 1135Hz Doppler shifts for training. Trained neural networks should generalize well to the channels \textbf{including but not just} those channels, which are just examples to show the generalization results. Details of the Doppler spectra for test are shown in Appendix~\ref{Appendix}. 
\begin{figure*}
    \centering
    \subfloat[NMSE results of InterpolateNet \label{Doppler_InterpolateNet}]{%
        \includegraphics[width=0.5\textwidth]{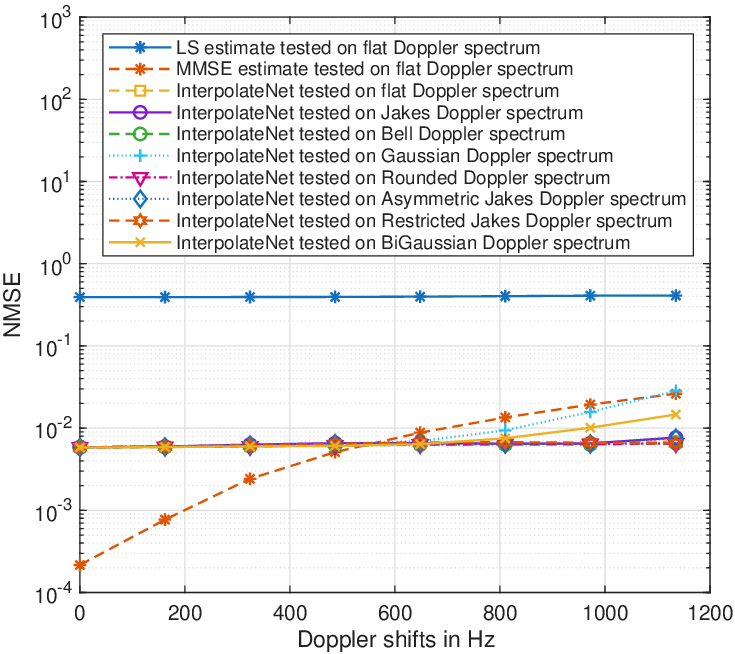}}
    \hfill
    \subfloat[NMSE results of Channelformer \label{Doopler_Channelformer}]{%
        \includegraphics[width=0.5\textwidth]{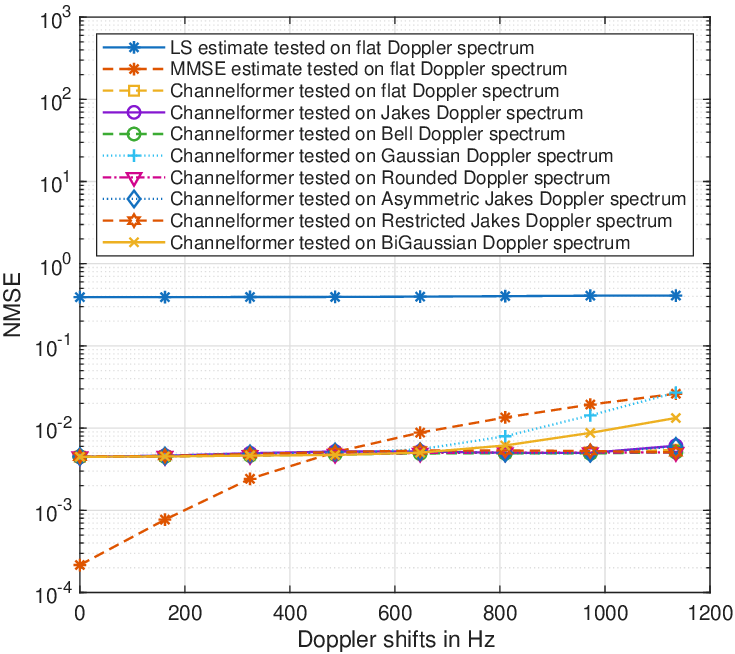}}
  \caption{NMSE results for the offline-trained InterpolateNet and Channelformer tested with different Doppler spectra, Doppler shifts and a fixed SNR of 20dB. The CE channel is set as default PDP. } 
  \label{Generalization to different Doppler}
\end{figure*}

Fig.~\ref{Doppler_InterpolateNet} provides the NMSE results for trained InterpolateNet tested on Doppler shifts from 0Hz to 1135Hz when $f_d$ has a Doppler spectrum of flat, Jakes, Bell, Gaussian, Round, Asymmetric Jakes and Restricted Jakes spectrum. For Doppler shifts from 0Hz to 1135Hz, both InterpolateNet and Channelformer achieve a flat NMSE performance for flat, Jakes, Bell, Round, Asymmetric Jakes, Restricted Jakes spectrum across the whole Doppler shift range. Compared to NMSE values at the Doppler shift of 0Hz, InterpolateNet increase NMSE values by 0.0088 on Gaussian spectrum at the Doppler shift of 1135Hz, while Channelformer increase NMSE values by 0.0087 on Gaussian spectrum. The LS estimate and the MMSE method have NMSE values from 0.39 to 0.40 and from 0.0002 to 0.0262, respectively. Because the MMSE method cooperates with bilinear interpolation for time-domain interpolation, it results in worse time-domain interpolation performance. Therefore, neural network methods can outperform the MMSE methods in a high Doppler shift range. Except for Gaussian spectrum, both InterpolateNet and Channelformer achieve almost identical MSE performance across these unseen Doppler spectra and Doppler shifts from 0Hz to 1135Hz, only with a slight variation of 0.00072 for InterpolateNet and 0.00065 for Channelformer at Doppler shift of 1135Hz. With the Gaussian spectrum, InterpolateNet achieves a NMSE value of 0.0282 and 0.0125 at Doppler shift of 1135Hz, respectively. Channelformer has a NMSE value of 0.0271 and 0.0112 for each, which also show a slight variation shown in Fig.~\ref{Generalization to different Doppler}. This degradation is due to that unlike other Doppler spectra, the Gaussian Doppler spectra in MATLAB are not strictly truncated to \([-f_d,f_d]\), and their out-of-band components lead to this, especially when Doppler shifts is very high. 
\subsection{Validation of preservation of generalization to PDPs}
\label{PDP_MSE}
After demonstrating the generalization to Doppler shifts and Doppler spectrum, we evaluate if this proposed design criteria will damage the design criteria for generalization to PDPs proposed in \cite{luan2025achieving}, while time correlation and frequency correlation are separable. 
\begin{figure*}
    \centering
    \subfloat[NMSE results for the offline-trained InterpolateNet and Channelformer tested on example channels given in Table.~\ref{customized channels} with a fixed SNR of 20dB. Each $f_{d,m}$ has a randomly assigned Doppler spectrum and the maximum mobile speed is 70km/h. \label{PDP}]{%
        \includegraphics[width=1.0\textwidth]{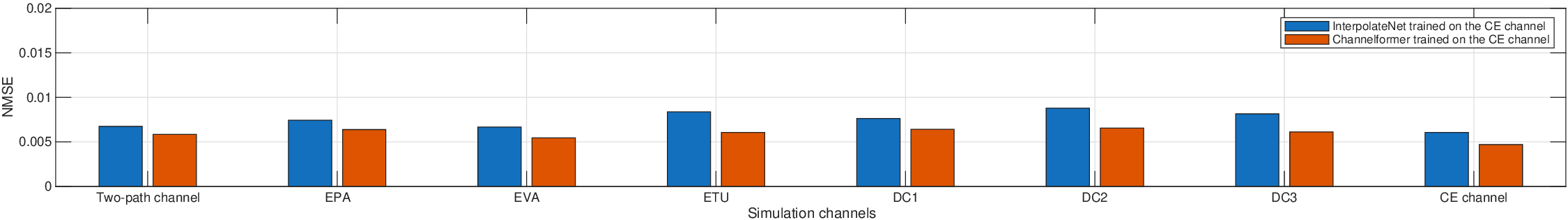}}
    \\
    \subfloat[Offline-trained InterpolateNet tested on randomly generated channels with a fixed SNR of 20dB, maximum mobile speed of 100km/h and each $f_{d,m}$ has a randomly assigned Doppler spectrum. \label{InterpolateNet_random}]{%
        \includegraphics[width=1.0\textwidth]{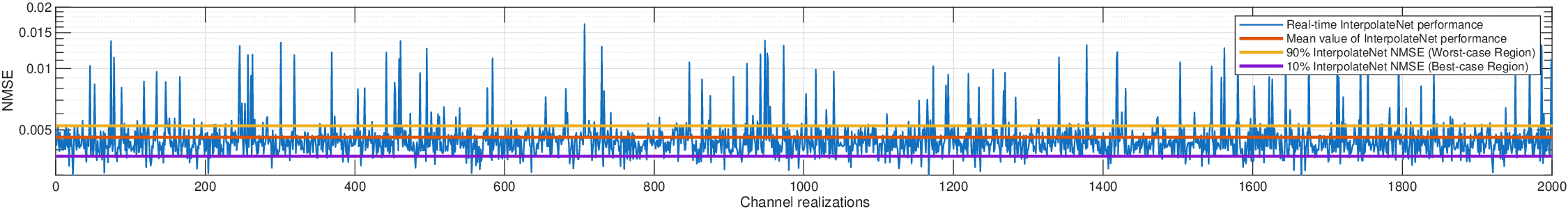}}
    \\
    \subfloat[Offline-trained Channelformer tested on randomly generated channels with a fixed SNR of 20dB, maximum mobile speed of 100km/h and each $f_{d,m}$ has a randomly assigned Doppler spectrum. \label{Channelformer_random}]{%
        \includegraphics[width=1.0\textwidth]{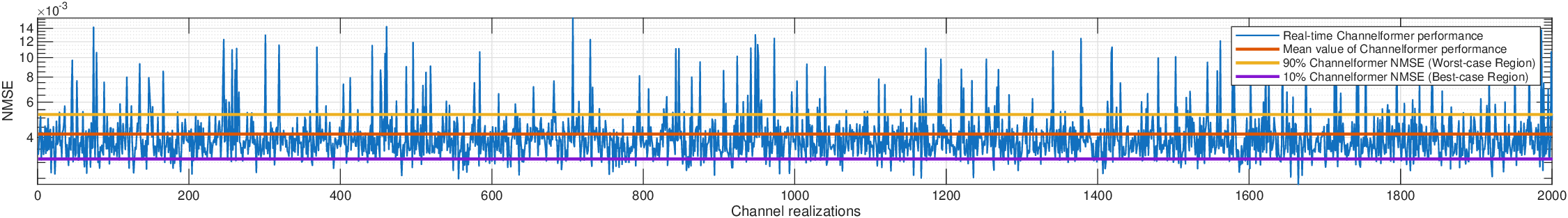}}
  \caption{Validation of generalization to different PDPs. Offline-trained InterpolateNet and Chnanelformer are tested on a randomly generated channel with a fixed SNR of 20dB and each path has a randomly assigned Doppler spectrum. }
  \label{Validation of generalization}
\end{figure*}

Fig.~\ref{PDP} provides the generalization property of offline-trained neural networks to different PDPs. For each path, the corresponding Doppler shifts have a randomly assigned Doppler spectrum. From the simulation results on these example channels, both InterpolateNet and Channelformer achieve almost identical NMSE values across these unseen PDPs. Fig.~\ref{InterpolateNet_random} and Fig.~\ref{Channelformer_random} show the NMSE values when trained InterpolateNet and Channelformer are tested on randomly generated channels and each NMSE point is averaged by 50 slots. Each channel realization has a path number $M$ randomly selected from 1 to 15, and the path delays are confined within a maximum delay spread of the duration of CP to emulate practical propagation characteristics. Up to three early paths are uniformly distributed within the first 700ns, subject to a minimum inter-path spacing constraint to prevent unrealistic clustering. The remaining paths are uniformly distributed over the interval from 700ns to the duration of CP, resulting in a sparser late-arrival structure. Each $a_m$ follows an exponential distribution to capture the typical power decay behavior of multipath channels. After power normalization, the strongest path is enforced within the early components. For offline-trained InterpolateNet and Channelformer, the mean value is 0.0046 and 0.0042 respectively; the upper boundary of the best-performing 10\% and the lower boundary of the worst-performing 10\% are 0.0037 and 0.0052 for InterpolateNet, 0.0030 and 0.0052 for Channelformer. These results demonstrate that the proposed design criteria do not compromise generalization to channels with unseen PDPs and remain robust even when the assumption of separable time–frequency channel correlations in (\ref{r_t}) is relaxed. 
\subsection{Ablation experiments: Validation of the proposed design criteria}
\label{Simulation_ablation}
As shown in Section.~\ref{Doppler_MSE}, trained neural networks exhibit strong generalization to Doppler by following the proposed design criteria. Simultaneously, we intentionally violate this design criteria for training to observe NMSE degradation, and this ablation experiment is used to understand the contribution of our proposal. 
\begin{figure*}[htbp]
\centerline{\includegraphics[width=0.85\textwidth]{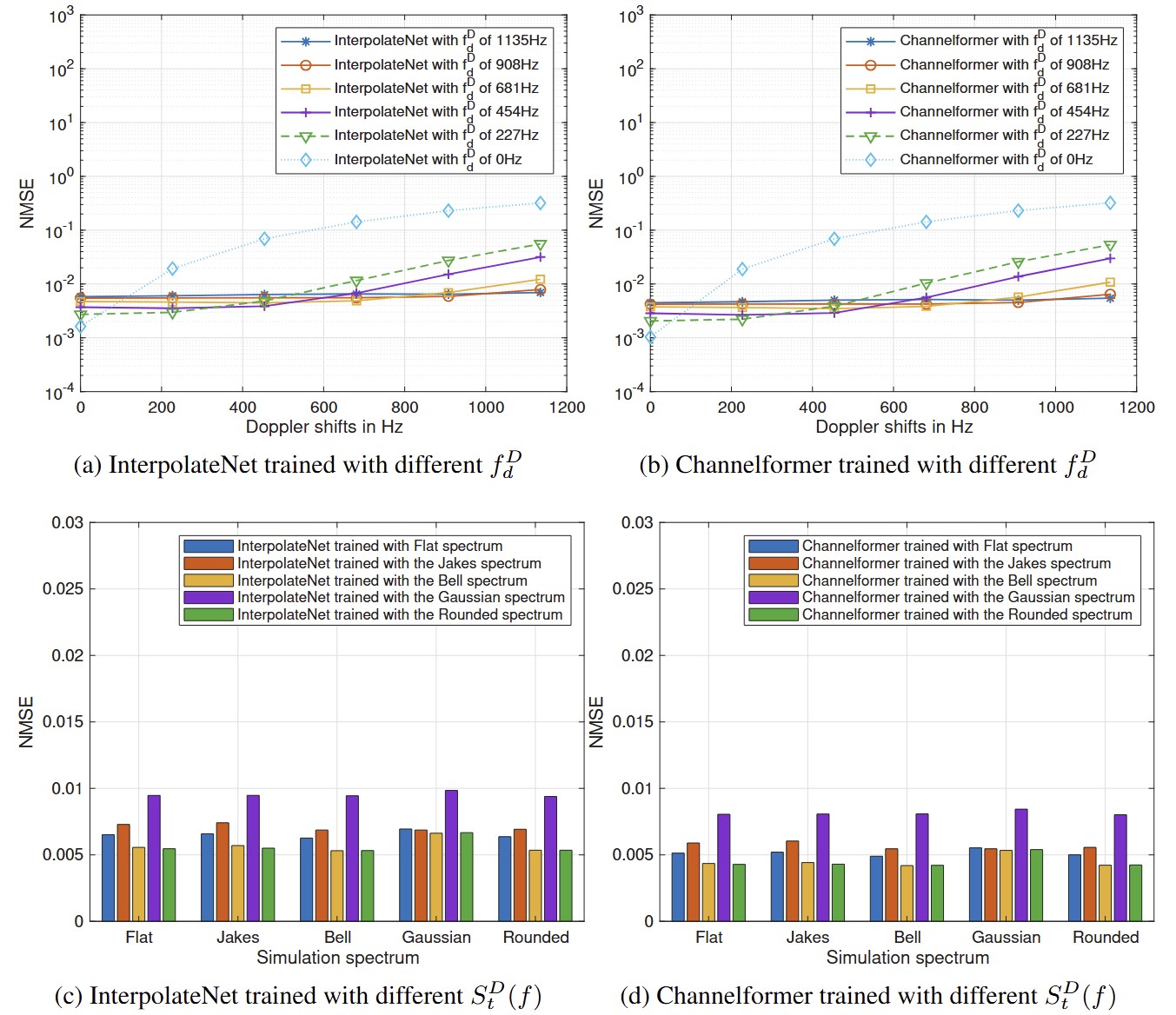}}
\caption{NMSE results for InterpolateNet and Channelformer trained with different $f_d^{D}$ and $S_t^{D}(f)$ and tested with different Doppler spectra, Doppler shifts and a fixed SNR of 20dB. The CE channel is set as default PDP. } 
\end{figure*}

Fig.~3a and Fig.~3b provide the NMSE values for InterpolateNets and Channelformers trained with different single $f_d^{D}$ and tested on Doppler shifts from 0Hz to 1135Hz with $S_t^{D}(f) = S_t^{A}(f)$ (flat spectrum). These curves increase as the Doppler shift increases. Neural networks trained with Doppler shift of 1135Hz achieve almost identical NMSE across these Doppler shifts. For example, InterpolateNet and Channelformer only have a slight variation from 0.0058 to 0.0068 and from 0.0045 to 0.0054 respectively. However, when the unseen $f_d^{A}$ are larger than the $f_d^{D}$ used for training, trained neural networks exhibit increased NMSE on these unseen $f_d^{A}$. For example, for Channelformer trained with $f_d$ = 681Hz, it has NMSE values of 0.0038, 0.0057 and 0.0108 at the maximum Doppler shifts of 681Hz, 908Hz and 1135Hz. Especially, for neural networks trained with lower Doppler shifts, they perform better on low Doppler shifts and degrade significantly on high Doppler range. For example, InterpolateNet trained with $f_d$ of 0Hz achieves the best NMSE among these trained InterpolateNets but exhibits the most severe increase on other unseen Doppler shifts. It has a NMSE of 0.0016 at 0Hz while the corresponding NMSE value of 0.3203 at 1135Hz, which indicates random prediction at high Doppler shifts. The increase in NMSE is mainly down to $f_d^{D} < f_{d}^{A}$, while the NMSE values are stable when $f_{d}^{A} \leq f_d^{D}$. 

Fig.~3c and Fig.~3d provide the NMSE values for InterpolateNets and Channelformers trained with different Doppler spectrum with $f_d^{D}$ = 640Hz, and then tested on these spectra with $f_d^{A}$ = 640Hz. InterpolateNet and Channelformer trained with flat spectrum has only slight NMSE variation of 0.000168 and 0.000143 among these spectra, which is the smallest variation compared to neural networks trained with other candidate spectrum. It should be noted that only the variation caused by channel mismatch is evaluated to show the generalization property. The aim is to ensure robustness against Doppler mismatch rather than to achieve the minimum estimate error among these spectra. From these simulation results, neural networks trained with flat spectrum and $f_d^{A} \leq f_d^{D}$ achieve good generalization to unseen Doppler spectra and Doppler shifts, while violating the proposed design criteria leads to an increase in NMSE. 
\section{Conclusion}
\label{Conclusion}
We propose design criteria to generate the training data for channel estimation neural networks that achieve strong generalization to unseen Doppler shifts and Doppler spectra. It ensures trained neural networks perform time-domain interpolation well across new and previously unseen channels, which retain the superior performance for practical implementation of channel estimation neural networks. Moreover, the proposed design criteria is applicable for most neural networks because no specific neural architecture is required. From the simulation results, trained neural networks achieve robust performance across different Doppler shifts and Doppler spectra. Moreover, the proposed design criteria do not damage the generalization to channels with unseen power delay profiles as shown in the simulation results. 
\subsection*{AI use statement}
This work used generative AI tools for assist with translation. We have not used generative AI tools to generate synthetic data sets, help develop theoretical models or conceptual frameworks, formulate mathematical claims, provide critical ingredients for proving mathematical claims, assist in the writing of proofs, propose or refine hypotheses, design or provide feedback on research methodology or experiments, implement methods, support qualitative and thematic data analysis, interpret results tasks, and clean and reformat dataset is not applicable to this work. In addition, we used generative AI tools to edit a research paper to improve readability. We use google scholar to generate the BibTeX for citations. We reviewed all AI-assisted work. We take responsibility for the final content of this work, including text, claims, or artifacts produced with the aid of generative AI. 
\bibliography{iclr2027_conference}
\bibliographystyle{iclr2027_conference}

\appendix
\section{Appendix}
\label{Appendix}
Assuming that the time correlation are same across path gains and the path gains are mutually uncorrelated wide-sense stationary complex Gaussian processes, we have 
\begin{equation}
r_{t, m}(l-l') = 
\mathbb{E}_{f_{d,m}}\!\left\{e^{j2\pi f_{d,m}T_o(l-l')}\right\}
= r_t(l-l'). 
\end{equation}
Then the channel correlation through both time domain and frequency domain is given by 
\begin{figure*}
\begin{equation}
\mathbb{E}\!\left\{
H(k,l)H^{*}(k',l')
\right\} = \frac{1}{N_f}\sum_{m=0}^{M-1} A_m r_{t, m}(l-l')
   e^{-j2\pi \frac{(k-k')\tau_m}{N_f}} = \left\|\mathbf p\right\|_1r_t(l-l')r_f(k-k'), 
\label{r_t} 
\end{equation}
\end{figure*}
where $A_m = \mathbb{E}\!\left\{a_{m}a_{m}^{*}\right\}$. Path gains $\{a_m\}_{m=0}^{M-1}$ are mutually independent. Thus, both $r_t, r_f$ are separable in channel correlation designs. As channel correlation $\mathbf{R_{HH}} = \mathbf{D}\Lambda\mathbf{D}^{H}$, for the $i$th diagonal element of its eigenvalue the corresponding spectrum is given by 
\begin{equation}
    S_{i}(f) = \mathbf{p}(i)S_{t}(f) + \sigma^{2}, 
\end{equation}
the noise item is introduced by the input of LS estimate and then this spectrum is normalized by $\sigma^{2}$ to obtain 
\begin{equation}
    S_{i}(f) = \frac{\mathbf{p}(i)}{\sigma^{2}}S_{t}(f) + 1. 
\end{equation}
Under the anti-causal assumption, we apply spectral factorization to this normalized power spectrum for one-side $M_i(f)$ that satisfy $\left\|M_i(f)\right\|_1^{2} = \frac{\mathbf{p}(i)}{\sigma^{2}}S_{t}(f) + 1$. Then convert $M_i(f)$ back by inverse Fourier transform to obtain $\Gamma_{i}$, we have the filter taps of $M_i(f)$ as 
\begin{equation}
    M_i(f) = \sum_{n = 0}^{\infty} \Gamma_{i}(n)e^{-jn2\pi f}, 
\label{M_i}
\end{equation}
To compute all taps $\Gamma_{i}$, it can be obtained by  
{\footnotesize
\begin{equation}
    M_i(f) = \exp\left\{0.5\int_{-0.5}^{0.5} \ln S_i(u) \, du + \sum_{n = 1}^{\infty} \int_{-0.5}^{0.5} \mathrm{ln}S_i(v)e^{jn2\pi (v-f)} \, dv\right\}. 
\end{equation}}
Then apply the inverse Fourier transform to obtain exact values. In this paper, $\Gamma_{i}(0)$ is calculated as 
\begin{equation}
    \Gamma_i(0) = \exp\left\{\frac{1}{2}\int_{-0.5}^{0.5}\ln\left(1+\frac{\mathbf{p}(i)S_t(f)}{\sigma^{2}}\right)\,df \right\}. 
\label{Gamma}
\end{equation}

Assume that both the designed Doppler spectrum and the environment Doppler spectrum with maximum Doppler shift of $f_d$ ($[-f_d, f_d] \subseteq [-0.5, 0.5]$ for DTFT normalized range) and satisfy
\begin{equation}
    \int_{-0.5}^{0.5} S_t^{A}(f)\,\mathrm{d}f = \int_{-0.5}^{0.5} S_t^{D}(f)\,\mathrm{d}f = 1.
\end{equation}
The mismatch error $\epsilon_{t, i}\!\left(S_t^{A};S_t^{D}\right)$ and the matched MSE are given by 
{\footnotesize
\begin{align}
\epsilon_{t, i}\!\left(S_t^{A};S_t^{D}\right) &= \int_{-0.5}^{0.5}\mathbf{p}(i)S_t^{A}(f)\left\|1-\mathbf{w}_i^{D}(f)\right\|_1^2+\sigma^2\left\|\mathbf{w}_i^{D}(f)\right\|_1^2\,df, \\
\epsilon_{t, i}\!\left(S_t^{D};S_t^{D}\right) &= \int_{-0.5}^{0.5}\mathbf{p}(i)S_t^{D}(f)\left\|1-\mathbf{w}_i^{D}(f)\right\|_1^2 + \sigma^2\left\|\mathbf{w}_i^{D}(f)\right\|_1^2 \, df. 
\end{align}}
Therefore, the total mismatch error is given by
\begin{equation}
\Delta\epsilon_t = \frac{1}{N_f}\sum_{i=1}^{N_f}\mathbf{p}(i)
\int_{-0.5}^{0.5}
\left(
S_t^{A}(f)-S_t^{D}(f)
\right)
\left\|1-\mathbf{w}_i^{D}(f)\right\|_1^2
\mathrm{d}f.
\end{equation}

The deployed Doppler spectra are described as follows with the constraint of $\left\|f\right\|_1\leq f_d$. The Rounded Doppler spectrum has the theoretical formula of 
\begin{equation}
    S(f)=\frac{1}{1.168f_d}\left[1-1.72\left(\frac{f}{f_d}\right)^2+0.785\left(\frac{f}{f_d}\right)^4\right]. 
\end{equation}
The Bell Doppler spectrum has the theoretical formula of 
\begin{equation}
S(f)=\frac{3}{2 \mathrm{tan}^{-1}(3) f_d}\frac{1}{1+9\left(\frac{f}{f_d}\right)^2},
\end{equation}
The Asymmetric Jakes Doppler spectrum is 
{\small
\begin{equation}
    S(f)=\frac{A_f}{\pi f_d\sqrt{1-\left(f/f_d\right)^2}},\ f_{\min}\leq f\leq f_{\max}, 
\end{equation}}
where $A_f = \pi\left[\sin^{-1}\left(\frac{f_{\max}}{f_d}\right)-\sin^{-1}\left(\frac{f_{\min}}{f_d}\right)\right]^{-1}$ and $f_{\min}$ = 0, $f_{\max} = f_d$. 

The Restricted Jakes Doppler spectrum has the theoretical formula of 
\begin{equation}
    S(f)=\frac{A_r}{\pi f_d\sqrt{1-\left(f/f_d\right)^2}},\quad 0\leq f_{\min}\leq \left\|f\right\|_1\leq f_{\max}\leq f_d, 
\end{equation}
where $A_r=\frac{\pi}{2}\left[\sin^{-1}\left(\frac{f_{\max}}{f_d}\right)-\sin^{-1}\left(\frac{f_{\min}}{f_d}\right)\right]^{-1}$, $f_{\min} = 0.2f_d$, $f_{\max} = 0.8f_d$. The Gaussian Doppler spectrum is 
\begin{equation}
    S_G(f)=\frac{1}{\sqrt{2\pi\sigma^2}}e^{-\frac{f^{2}}{2\sigma^2}}, \sigma = \frac{1}{\sqrt{2}}f_d. 
\end{equation}

\end{document}

%% file: math_commands.tex
\usepackage{amsmath,amsfonts,bm}

\def\eqref#1{equation~\ref{#1}}

\def\1{\bm{1}}

\DeclareMathAlphabet{\mathsfit}{\encodingdefault}{\sfdefault}{m}{sl}
\SetMathAlphabet{\mathsfit}{bold}{\encodingdefault}{\sfdefault}{bx}{n}

